\documentclass[aps, prb, reprint, superscriptaddress, floatfix, showkeys]{revtex4-2}

\usepackage[utf8]{inputenc}
\usepackage[T1]{fontenc}
\usepackage{amsmath, amssymb, amsfonts, amsthm}
\usepackage{graphicx}
\usepackage{bm} 
\usepackage{xcolor}
\usepackage{booktabs, multirow, soul}
\usepackage[normalem]{ulem}
\usepackage{url}
\usepackage{mathrsfs}
\usepackage{kotex}
\usepackage{textcomp}
\usepackage{mathtools}
\usepackage{romannum}

\usepackage{algorithm}
\usepackage{algpseudocode}
\usepackage{listings}
\usepackage{lineno}

\usepackage[colorlinks=true, citecolor=blue, urlcolor=blue, linkcolor=blue]{hyperref}

\begin{document}

\pagenumbering{arabic}

\title{Optimized control protocols for stable skyrmion creation using deep reinforcement learning}

\author{Ji Seok Song}
\affiliation{Graduate School of Quantum Science and Technology, Korea Advanced Institute of Science and Technology, Daejeon 34141, Republic of Korea}

\author{Se Kwon Kim}
\affiliation{Graduate School of Quantum Science and Technology, Korea Advanced Institute of Science and Technology, Daejeon 34141, Republic of Korea}
\affiliation{Department of Physics, Korea Advanced Institute of Science and Technology, Daejeon 34141, Republic of Korea}


\author{Kyoung-Min Kim}
\email{kyoungmin.kim@apctp.org}
\affiliation{Asia Pacific Center for Theoretical Physics, Pohang 37673, Republic of Korea}
\affiliation{Department of Physics, Pohang University of Science and Technology, Pohang 37673, Republic of Korea}

\date{\today}

\begin{abstract}
Generating stable magnetic skyrmions is essential for the practical application of skyrmion-based spintronic devices in thermally agitating environments. Here, we present a deep reinforcement learning (DRL) approach to identify advanced dynamic magnetic-field-temperature paths that create skyrmions with enhanced thermal stability. The trained DRL agent discovers an optimized field-temperature path that achieves a higher success rate for skyrmion formation in Fe\textsubscript{3}GeTe\textsubscript{2} monolayers compared to previous fixed-temperature field sweeps. Additionally, the generated skyrmions exhibit longer lifetimes due to their isotropic shape and equilibrium size, both of which place them near a local energy minimum and thereby hinder annihilation. We demonstrate that these advancements stem from the targeted minimization of the dissipated work, which ensures that the driven skyrmion states remain close to their equilibrium distributions by upper-bounding the Kullback–Leibler divergence. Our findings suggest that a physics-informed DRL framework streamlines the identification of optimized protocols for skyrmion creation.
\end{abstract}


\maketitle


\section{Introduction}\label{sec1}

Magnetic skyrmions—topological spin textures characterized by non-zero integer topological charges—are anticipated to be essential building blocks for future applications in spintronics and unconventional computing techniques \cite{fert2017magnetic, chen2019sky, Song2020, Marrows2024}. To realize these applications, developing methods for the controlled generation of isolated skyrmions is crucial---particularly for logic gates~\cite{Zhang2015logic, Chauwin2019reversible}, racetrack memories~\cite{Tomasello2014racetrack, Woo2018deterministic}, and neuromorphic synapses~\cite{Song2020, Marrows2024}---and significant progress has been made in recent years \cite{Zhang_2020}. 
However, the inherent stochasticity of the nucleation process within these methods often limits the success rate, rendering their creation probabilistic rather than fully reliable \cite{Zazvorka2019, doi:10.1021/acs.nanolett.7b00649}. Furthermore, thermal fluctuations undermine the efficacy of topological protection, forcing isolated skyrmions into transient states and significantly shortening their functional lifetimes at finite temperatures \cite{wild2017entropy, Hagemeister2015}. Furthermore, magnetic parameters such as single-ion anisotropy often exhibit strong temperature dependence, which can lead to the failure of skyrmion creation during transitions between the start and end points in the parameter space, as observed in Fe\textsubscript{3}GeTe\textsubscript{2} (FGT) \cite{birch2022history}. Consequently, achieving both deterministic formation and long-term thermal stability of isolated skyrmions remains a critical hurdle for the development of robust skyrmion-based technologies \cite{Cortes-Ortuno2017-ew}.

\begin{figure}[t!]
    \centering
     \includegraphics[width=\linewidth]{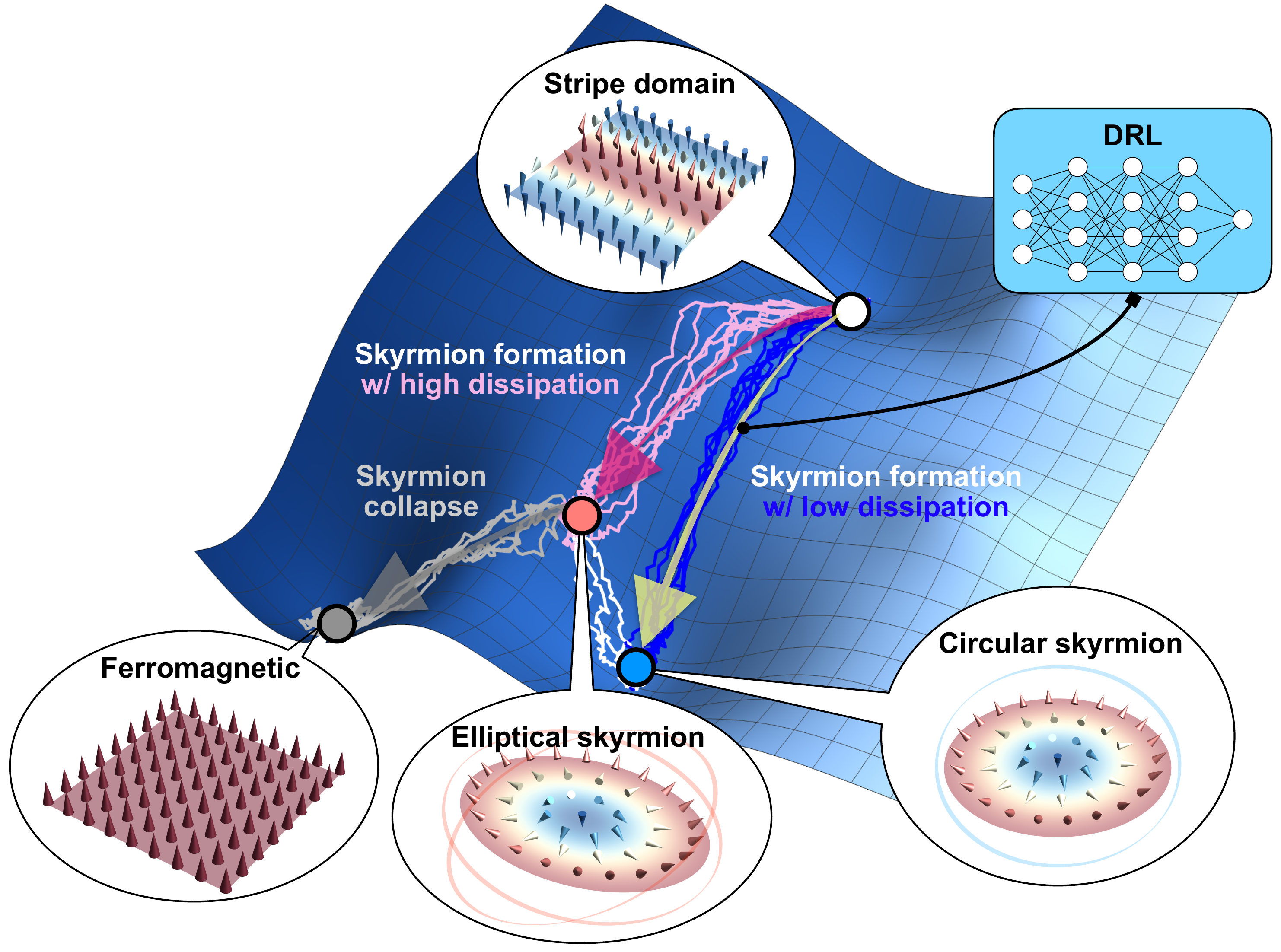}
    \caption{
    \textbf{Schematic of the DRL framework for identifying optimized skyrmion-creation protocols.} The trained DRL agent generates a time-dependent protocol for the external magnetic fields and temperature (indicated by the yellow arrow). This control drives the system from an initial stripe domain state to an isolated skyrmion state. By minimizing dissipation, the protocol yields a fully relaxed, circular skyrmion positioned near a local energy minimum. This energetic stability effectively reduces the probability of the skyrmion collapsing into the ferromagnetic ground state during post-protocol relaxation (indicated by the gray arrow), thereby ensuring enhanced longevity. In contrast, the conventional protocol (indicated by the magenta arrow) induces high dissipation, resulting in elliptical skyrmions with higher energy. Such states reside far from the local energy minimum, exhibiting reduced energetic stability and significantly shorter lifetimes. 
    }
    \label{fig1}
\end{figure}

In this study, we present a deep reinforcement learning (DRL) approach to identify high-fidelity skyrmion creation protocols that simultaneously enhance thermal stability. Our DRL agents are trained to generate dynamic control protocols for the magnetic field and temperature within micromagnetic simulations of FGT monolayers, facilitating skyrmion nucleation from an initial stripe domain configuration [Fig.~\ref{fig1}]. We design a cost function for the training as a sum of the topological charge and the dissipated work, which are determined by the final spin configuration and the magnetic field and temperature profiles, respectively. 
Compared to the previous cost function based solely on the final spin configuration \cite{Wang_2025}, our scheme exploits thermodynamic constraints, thereby enhancing the thermal stability of the created skyrmions. By optimizing this cost function, the agents discover advanced control protocols that achieve a significantly higher success rate for thermal-activation-based nucleation compared to conventional fixed-temperature field sweeps \cite{moon2021universal}. Moreover, through the minimization of dissipated work, the learned protocols produce skyrmions that retain an isotropic shape and equilibrium size, thereby exhibiting enhanced longevity compared to those generated under highly dissipative control schemes. We attribute this enhanced performance to the minimized dissipation, which ensures that the generated skyrmions are situated closer to their equilibrium states, thereby providing superior thermal stability and functional longevity.

\section{Results}\label{sec2}

Our micromagnetic simulations are performed on an FGT monolayer. This material demonstrates skyrmion stabilization due to its substantial perpendicular magnetic anisotropy and interfacial Dzyaloshinskii-Moriya interaction (DMI) arising from the broken inversion symmetry at the substrate interface \cite{doi:10.1126/sciadv.abb5157, birch2022history, Wu2020}.
The magnetic energy density used in the simulation is given by
\begin{equation} \label{eq:free_energy}
\begin{split}
    \mathcal{F} = & A (\nabla \bm{m})^2 - K m_z^2 + D \Big[m_z(\nabla\cdot \bm{m})-(\bm{m}\cdot\nabla)m_z\Big] \\
    & - \mu_0 M_s \bm{m} \cdot \bm{H} - \frac{\mu_0}{2} M_s \bm{m} \cdot \bm{H}_{\text{demag}},
\end{split}
\raisetag{15pt}
\end{equation}
where $\bm{m} = (m_x,m_y,m_z)$ is the unit magnetization vector and $M_s$ is the saturation magnetization at temperature $T$. The parameters $A$, $K$, and $D$ denote the ferromagnetic (FM) exchange stiffness, the uniaxial anisotropy constant at temperature $T$, and the interfacial DMI strength, respectively. In FGT, both \(M_{s}\) and \(K\) exhibit a strong temperature dependence \cite{liu2024anomalous, liu2025exotic}, which significantly influences skyrmion formation \cite{birch2022history}. Consequently, we adopted the temperature-dependent functions from Ref.~\citenum{liu2025exotic} for \(M_{s}\) and \(K\). The vector $\bm{H}=(H_x,H_y,H_z)$ is the externally applied magnetic field, and $\bm{H}_{\text{demag}}$ is the demagnetizing field. To incorporate thermal effects, we performed micromagnetic simulations using the stochastic Landau-Lifshitz-Gilbert (LLG) equation \cite{leliaert2017adaptively} via the \texttt{MuMax3} software package \cite{10.1063/1.4899186}. Periodic boundary conditions were imposed in the in-plane directions to minimize finite-size effects. Specific magnetic parameters---including $A$, $K$, $D$, $M_s$, and the Gilbert damping coefficient---as well as geometry details are provided in Supplementary Note~3.

\begin{figure*}[t!]
    \centering
    \includegraphics[width=1\linewidth]{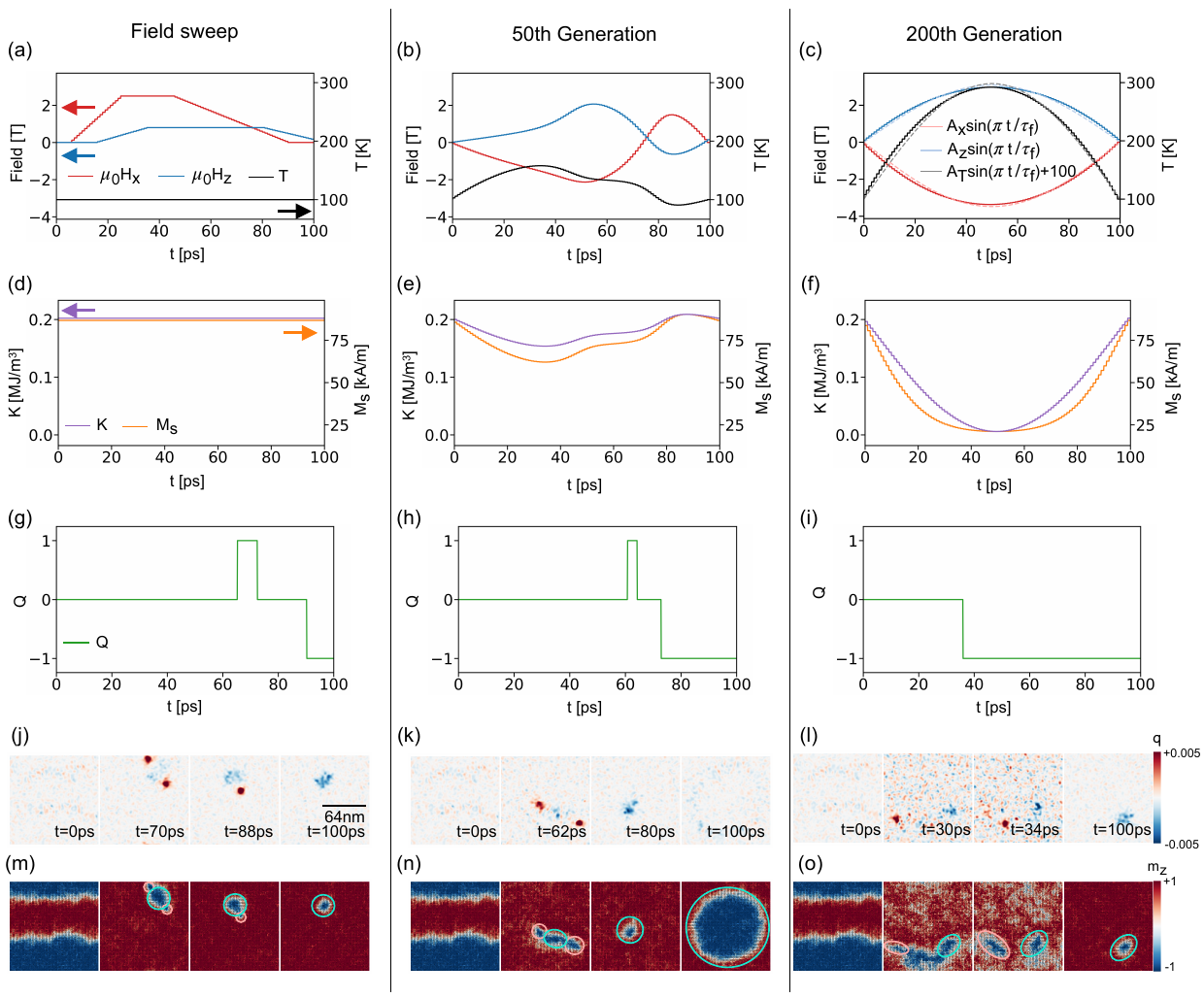}
    \caption{\textbf{Comparison of control protocols and corresponding magnetization dynamics in Fe\textsubscript{3}GeTe\textsubscript{2} (FGT).} The figure is organized into three columns, representing: a conventional field-sweep protocol (left), a protocol generated by a 50th generation DRL agent (center), and a protocol from a 200th generation DRL agent (right). (a)--(c) Time-evolution of the control parameters, including temperature ($T$) and magnetic field components ($\mu_0 H_x, \mu_0H_z$), for the respective protocols. Dashed lines represent sinusoidal fitting functions with the fitted parameters: $A_x = -3.48\mathrm{T}$, $A_z = 3.09\mathrm{T}$, $A_T = 198.26\mathrm{K}$, and $\tau_f = 100~\mathrm{ps}$. (d)--(f) Corresponding time-dependence of the Hamiltonian parameters for FGT, specifically the uniaxial anisotropy ($K$) and saturation magnetization ($M_s$). 
    (g)--(i) Evolution of the topological charge $Q$ during exemplary skyrmion creation processes under the protocols outlined in (a)--(c). 
    (j)--(l) Snapshots of the topological charge density $q(x,y)$ corresponding to (g)--(i), respectively, with specific timestamps labeled in each panel. 
    Positive (red) and negative (blue) regions mark the charge densities corresponding to an antiskyrmion and a skyrmion, respectively.
    (m)--(o) Spin configurations corresponding to (j)--(l), respectively, where local arrows denote the in-plane magnetization direction at each spin site.
    The sky-blue circle denotes a skyrmion ($Q = -1$), while the pink circle denotes an antiskyrmion ($Q = +1$).
    }
    \label{fig2}
\end{figure*}


For comparison, we first benchmark the conventional fixed-temperature field sweep method \cite{moon2021universal}. Starting from an initial stripe domain state [Fig.~\ref{fig2}(m), first panel], an in-plane field $H_x$ is applied [Fig.~\ref{fig2}(a), red line; from 25 to 45 ps], which reduces the stripe width by canting the spins toward the in-plane direction. The subsequent application of an out-of-plane field $H_z$ [Fig.~\ref{fig2}(a), blue line; from 35 to 80 ps] causes the stripe to ``pinch off'', forming a precursor domain-like pattern [Fig.~\ref{fig2}(m), second panel]. Finally, the spin configuration is relaxed from 80 to 100 ps by gradually ramping $(\mu_0H_x, \mu_0H_z)$ toward the target values of $(0, 0.15\,\text{T})$. This final relaxation step leads to three distinct scenarios. First, an isolated skyrmion with $Q=-1$ is successfully created by the end of the simulation, accounting for $16.4\%$ of the cases out of 700 cases. 
Here, $Q = \iint q(x,y) \, dxdy$ represents the topological charge, defined via the local charge density $q(x,y)=\frac{1}{4\pi} \bm{m} \cdot (\partial_x \bm{m} \times \partial_y \bm{m})$. 
The intermediate state in this process displays one skyrmion and two antiskyrmions bound together [Figs.~\ref{fig2}(j) and (m), second panels]; it can alternatively accommodate a single skyrmion–antiskyrmion pair with a probability comparable.
The successive annihilation of these antiskyrmions [third panels] eventually leads to an isolated single skyrmion [fourth panels]—a sequential process clearly resolved in Fig.~\ref{fig2}(g) by the time evolution of the topological charge stepping through $Q=1$, $Q=0$, and finally $Q=-1$.
Second, an antiskyrmion with $Q = +1$ is nucleated in $7.4\%$ of the trials; however, this antiskyrmion is inherently unstable and eventually annihilates upon further relaxation \cite{Stability2023Potkina, koshibae2016theory, hoffmann2017antiskyrmions}, as will be elucidated later. 
Third, in $39.4\%$ of the cases, a single skyrmion may develop an elongated geometry and subsequently annihilates during relaxation.
The detailed spin dynamics corresponding to the latter two cases are provided in Supplemental Videos 1 and 2. Finally, the remaining $36.8\%$ of the cases result in unsuccessful nucleation, where the initial stripe domain simply fails to pinch off and persists throughout the entire protocol.

The DRL agent is designed to generate time-dependent trajectories for the external magnetic field and temperature within micromagnetic simulations, driving the transformation of an initial stripe domain configuration into skyrmions. During training, the agent's parameters are progressively optimized to minimize the following cost function:
\begin{equation}\label{eq:cost_function}
     \phi[H_x(t), H_z(t), T(t)] = ||Q_f| - 1| + k_s \frac{W_f}{k_BT_f}.
\end{equation}
In the first term, $Q_f\equiv Q(t=\tau_f)$ represents the topological charge at the final simulation time \(\tau_f\). By ensuring $ |Q_f| = 1$, the agents facilitate the transformation of an initial stripe domain configuration into a skyrmion state. In the second term, $W_f$ accounts for the dissipated work generated during the control trajectory, defined as: $W_f = \Delta E(\tau_f) + T_{f} \Delta S_{\text{res}}(\tau_f)$, where $\Delta E(\tau_f)$ is the change in internal magnetic energy and $T_{f}$ is the temperature at the final time, which is identical to that at the initial time. The term $\Delta S_{\text{res}}(\tau_f)$ quantifies the entropy production in the thermal reservoir:
\begin{equation}
\Delta S_\text{res}(\tau_f)=\int^{\tau_f}_{0} dt \int d\bm{r} \frac{M_{s}(T(t))}{T(t)} \, \mu_0 \, \dot{\bm{m}}(\bm{r}) \cdot \bm{H}_{\text{eff}}[\bm{m}].
\end{equation}
This formula is derived following Ref.~\citenum{bandopadhyay2017rotational}; see Supplementary Note~1 for the full derivation. Here, \(\bm{H}_\mathrm{eff}[\bm{m}] = -\frac{1}{\mu_0 M_s} \frac{\partial \mathcal{F}}{\partial \bm{m}}\) is the effective magnetic field derived from the magnetic energy density $\mathcal{F}$. The weighting factor $k_s=1.38 \times 10^{-8}$ controls the relative influence of the dissipated work against the topological charge within the cost function. It is worth noting that our definition of $W_f$ departs from the traditional formulation for isothermal processes, $W_f^\text{iso} = W - \Delta F$ \cite{kawai2007dissipation, vaikuntanathan2009dissipation}, where $W$ represents the work performed on the system and $\Delta F$ denotes the change in free energy. While $W_f$ and $W_f^\text{iso}$ are identical at a fixed temperature, temperature-varying control protocols necessitate the use of the generalized $W_f$. A rigorous derivation of this general form, along with a proof of its equivalence to $W_f^\text{iso}$ under isothermal conditions, is provided in Supplementary Note~2.

The inclusion of dissipated work in the cost function is intended to enhance the thermal stability of the generated skyrmions.
In general, driving a system via external control parameters---such as magnetic fields and temperature---typically results in a final state distribution that deviates from the thermal equilibrium distribution dictated by the final parameter values, even when the initial states are sampled from equilibrium. This behavior is a fundamental consequence of the second law of thermodynamics \cite{vaikuntanathan2009dissipation}. Consequently, the distribution of magnetic states produced in our simulations inevitably departs from the corresponding thermal equilibrium distribution. This deviation is closely linked to the thermodynamic lag \cite{vaikuntanathan2009dissipation} in the skyrmion formation process [see schematic illustration in Fig.~S2], and can be quantitatively measured by the Kullback-Leibler (KL) divergence:
\begin{equation} \label{eq:KLdivergence_E}
    D_\text{KL}\big(\rho||\rho_\text{eq}\big) = \int \rho(\mathcal{F}) \ln \frac{\rho(\mathcal{F})}{\rho_\text{eq}(\mathcal{F})} d\mathcal{F},
\end{equation}
where $\rho(\mathcal{F})$ and $\rho_\text{eq}(\mathcal{F})$ represent the simulated distribution of magnetic states as a function of $\mathcal{F}$ and the ideal thermal equilibrium distribution, respectively. Crucially, the KL divergence is upper-bounded by the dissipated work $W_f$ \cite{vaikuntanathan2009dissipation}:
\begin{equation} 
    D_\text{KL}\big(\rho || \rho_\text{eq}\big) \le \frac{W_f}{k_BT_f}.
\end{equation}
This relationship suggests that by minimizing $W_f$, $\rho(\mathcal{F})$ can be brought closer to $\rho_\text{eq}(\mathcal{F})$. In other words, suppressing the dissipated work favors the formation of skyrmion configurations that remain as close as possible to their local energy minima, given the constraints of a thermally agitating environment. This effective ``energy minimization'' can enhance the stability of skyrmions against thermal fluctuations.

In the training process, the agent's trajectory was initialized using the aforementioned fixed-temperature field sweep method and subsequently refined through successive optimization cycles. To account for stochasticity, the cost function $\phi$ was averaged over 10 independent trajectories, each subject to a different realization of the thermal field. This approach ensures that the agent optimizes the cost function across the thermal ensemble, rather than a single trajectory. Previous research indicates that optimizing the cost function efficiently requires prioritizing the target magnetic configuration---here, a unit topological charge---before minimizing dissipated work \cite{whitelam2023demon}. We find that an optimal value of $k_s = 1.38 \times 10^{-8}$ successfully implements this hierarchy, ensuring the unit topological charge is secured before work minimization dominates the later training stages. The optimization was executed via a genetic algorithm over 200 iterations to produce successive ``generations'' \cite{such2018deep}. Detailed implementation of the genetic algorithm is provided in Supplementary Notes~5 and 6, with schematics of the DRL agent architecture and the genetic training process presented in Figs.~S4 and S5, respectively.

Figures~\ref{fig2}(b)--(c) illustrate the external magnetic field and temperature control protocols developed by the 50th and 200th generations, representing intermediate and maturely trained versions, respectively. Both agents evolve strategies characterized by field magnitudes exceeding those of the field-sweep method and exploit substantial thermal elevations. We posit that the agents strategically leverage thermally-randomized spin configurations, induced by these temperature increases, to facilitate the formation of the in-plane winding textures necessary for skyrmion nucleation~\cite{Wang2020Thermal, Chen2022Thermal}. Furthermore, these elevated temperatures provide an additional advantage by reducing both the magnetic anisotropy and saturation magnetization, effectively lowering the energy barrier required to transition into the requisite in-plane winding states [Figs.~\ref{fig2}(e)--(f)] relative to the static-parameter landscape of the field-sweep method [Fig.~\ref{fig2}(d)]. We find that the control protocols at the 200th generation are well-fitted to sinusoidal forms, yielding $R^2 \approx 0.98$ in all cases, as shown in Fig.~\ref{fig2}(c).


By leveraging their developed conditions described above, the agents at the 50th and 200th generations successfully drive the transformation of the initial stripe domain state into isolated skyrmions. As illustrated in Figs.~\ref{fig2}(n)--(o), the agent-predicted trajectories follow a three-step nucleation process qualitatively similar to the field-sweep method: (\Romannum{1}) compression of the stripe width via an in-plane field; (\Romannum{2}) formation of bubble-like domains under the subsequent out-of-plane field, accompanied by the nucleation of skyrmions and antiskyrmions;
and (\Romannum{3}) relaxation toward the target field configuration, during which the antiskyrmion annihilates, resulting in an isolated skyrmion state. 
Their detailed processes, however, exhibit substantial discrepancies.
Specifically, while the intermediate state for the 50th-generation agent can accommodate either one skyrmion and two antiskyrmions [second panels of Figs.~\ref{fig2}(k) and (n)] or a single skyrmion–antiskyrmion pair (not shown), that for the 200th-generation agent exclusively consists of a single skyrmion–antiskyrmion pair [second panels of Figs.~\ref{fig2}(l) and (o)].
Furthermore, the stabilization of an isolated single skyrmion occurs significantly earlier with the 200th-generation agent than with the 50th-generation agent [Figs.~\ref{fig2}(h) and (i)].
The detailed spin dynamics corresponding to Figs.~\ref{fig2}(n)--(o) are provided in Supplemental Videos 3 and 4. Notably, skyrmions generated by the two protocols exhibit significant differences in size. Skyrmions generated at the 50th generation exhibit significant size dispersion, with areas of $A =4791 \pm 2140~\text{nm}^2$; for instance, the skyrmion presented in the fourth panel of Fig.~\ref{fig2}(n) possesses a size $A=7697~\text{nm}^2$, considerably exceeding the equilibrium value of $A=272~\text{nm}^2$. In contrast, skyrmions produced at the 200th generation closely match the equilibrium size ($A=282\pm 133~\text{nm}^2$), as exemplified by $A=240~\text{nm}^2$ shown in the fourth panel of Fig.~\ref{fig2}(o). Here, the area $A$ is defined as the area where the out-of-plane magnetization is antiparallel to the FM background; see Supplementary Note~4 for the detailed calculation of the skyrmion size. It is worth noting that the main advance of the DRL framework lies in its ability to reliably nucleate the same stable skyrmion state as the field-sweep method, rather than discovering a new or more stable one.

\begin{figure*}[t!]
    \centering
     \includegraphics[width=1\linewidth]{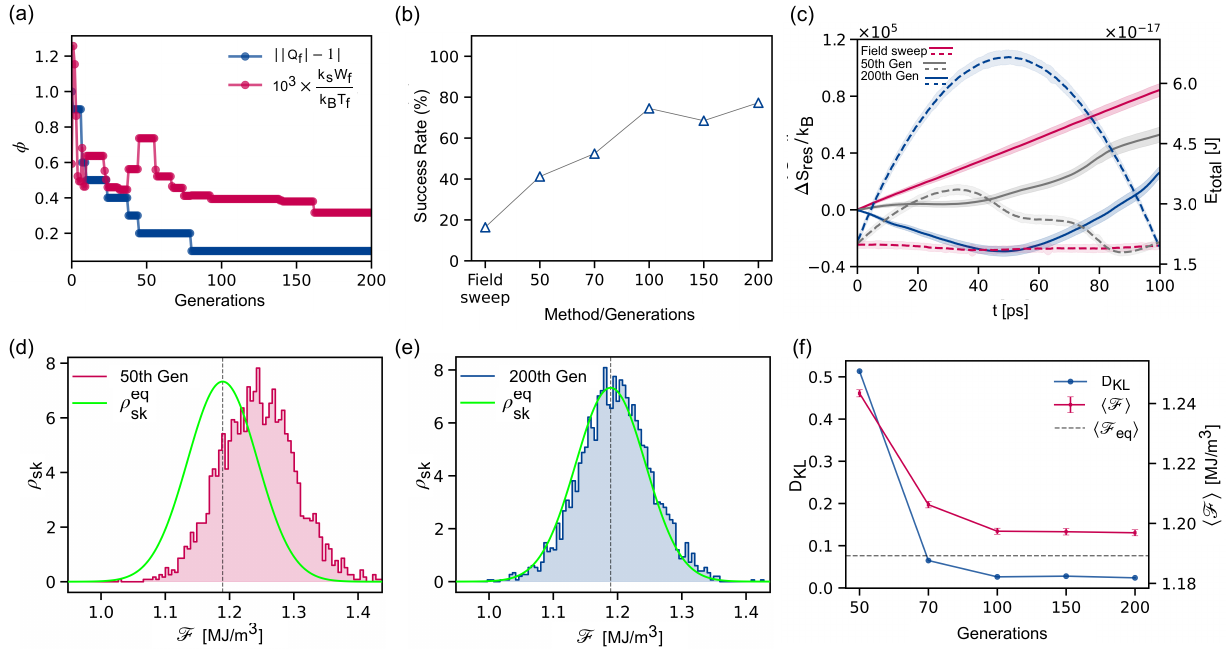}
    \caption{\textbf{Enhanced nucleation efficiency and thermal stability of skyrmions via DRL.} (a) Evolution of the two terms in the cost function $\phi$ over training epochs from generation 0 to 200. The first term represents $||Q_f|-1|$, while the second term corresponds to $k_s W_f/(k_B T_f)$; both terms are averaged over 10 independent trajectories. For visual clarity, the second term is scaled by a factor of $10^3$ to bring both contributions into a comparable numerical range. (b) Success rate of skyrmion formation for the conventional field-sweep protocol and for protocols generated at different DRL generations. 
    (c) Time evolution of the cumulative reservoir entropy production $\Delta S_{\mathrm{res}}/k_B$ (left axis, solid lines) and the total magnetic energy $E_{\mathrm{total}}$ (right axis, dashed lines) for each protocol. Shaded regions represent the standard deviation ($\pm 1\sigma$) from the mean.     
    (d)–(e) Probability density functions of the magnetic energy density $\mathcal{F}$ for instantaneous skyrmion states generated by the 50th and 200th generation DRL protocols, respectively (shaded areas). In each panel, the green lines represent the equilibrium distribution of $\mathcal{F}$ for the skyrmion state. (f) The left axis displays Kullback-Leibler (KL) divergence ($D_{\mathrm{KL}}$) of the 50th, 70th, 100th, 150th, and 200th DRL generations while the right axis shows the corresponding ensemble-averaged magnetic energy density $\langle \mathcal{F} \rangle$ of these instantaneous skyrmion states, together with the equilibrium value $\langle \mathcal{F}_{\mathrm{eq}} \rangle$ (dashed line). Error bars denote the standard error of the mean.
    }
    \label{fig3}
\end{figure*}

To gain deeper insight into the evolution of the DRL agent's capabilities, we examine the individual components of the cost function as a function of the training iteration. As illustrated in Fig.~\ref{fig3}(a), the deviation of the topological charge from unity---the first term defined in Eq.~\eqref{eq:cost_function}---exhibits a monotonic decrease. Given that a topological charge of unity signifies successful skyrmion nucleation, this trend reflects a consistent improvement in the agent's nucleation fidelity. Notably, the skyrmion creation success rate increased significantly throughout the training process, rising from 41.3\% out of 300 cases at 50th generation to 77.5\% by 200th generation out of 200 cases [Fig.~\ref{fig3}(b)]. This represents a marked improvement over the 16.4\% success rate achieved by the baseline fixed-temperature field sweep method. We attribute this enhanced performance to the agent's emergent strategy of ``borrowing'' thermal energy from the environment. As demonstrated in Fig.~\ref{fig3}(c), the control protocol for the 200th generation exhibits negative values of entropy production during the initial simulation stages. Within the initial period, the system absorbs energy from the environment, as governed by the relationship $\delta Q = -T \delta S_{\text{res}} = -\mu_0 M_s \delta t \int d\bm{r}\bm{H}_{\text{eff}} \cdot \dot{\bm{m}}$. Here, $\delta Q$ represents the heat exchange, where a positive value ($\delta Q > 0$) indicates energy absorption by the system, and $\delta S_{\text{res}}$ denotes the entropy change of the reservoir during the interval $\delta t$~\cite{bandopadhyay2017rotational}. 
The time evolution of the total magnetic energy confirms that in the 200th generation protocol steers the system to absorb thermal energy from the reservoir [dashed lines in Fig.~\ref{fig3}(c)]. Meanwhile, the energy absorption is not substantial in the 50th-generation protocol or completely absent in the field sweep method.
The absorbed thermal energy enables the system to surmount local energy barriers, thereby facilitating the nucleation of the skyrmion state with higher probability \cite{Wang2020Thermal, Chen2022Thermal, GARANIN2020165724}. This energy absorption mechanism stands in sharp contrast to the field-sweep method, which exhibits a monotonic increase in entropy production. Similar transient absorptions of thermal energy have been documented in other DRL studies on magnetic systems~\cite{whitelam2023demon}. Ultimately, the cumulative entropy production in the simulation becomes positive, resulting in the dissipation of energy to the environment as heat over the total duration [Fig.~\ref{fig3}(c)]. A minor dip in the success rate occurs between the 100th and 150th generations [Fig. \ref{fig3}(b)], likely due to statistical sampling fluctuations during training. For a full analysis, see Supplementary Note~7.

To delve deeper into the origin of the transient negative entropy production discussed above, we decompose the reservoir entropy production into the following form (the full derivation is provided in Supplementary Note~1): $\Delta S_\mathrm{res} = \Delta S_\mathrm{res}^{(\mathrm{det})} + \Delta S_\mathrm{res}^{(\mathrm{th})}$, where $\Delta S_\mathrm{res}^{(\mathrm{det})}$ and $\Delta S_\mathrm{res}^{(\mathrm{th})}$ are given by
\begin{equation}
\begin{aligned}
    \Delta S_\mathrm{res}^{(\mathrm{det})} &= \eta \mu_0\alpha \int dt \int d\bm{r} \;\frac{M_s}{T}\, |\bm{m}\times\bm{H}_\mathrm{eff}|^2, \\
    \Delta S_\mathrm{res}^{(\mathrm{th})} &= \eta \mu_0 \int dt \int d\bm{r} \; \frac{M_s}{T} \tilde{H}_\mathrm{th}^\mathrm{dmp}. 
\end{aligned} 
\label{eq:Sres_decomp}
\end{equation}
Here, $\eta = \gamma / (1+\alpha^2)$ is a gyromagnetic prefactor originating from the LLG equation, where $\gamma$ is the gyromagnetic ratio, $\alpha$ is the dimensionless Gilbert damping coefficient, and $\mu_0$ is the vacuum permeability.
The quantity $\tilde{H}_\mathrm{th}^\mathrm{dmp} = H_{\mathrm{th}, 1} - \alpha H_{\mathrm{th}, 2}$ represents the effective thermal drive for dissipation. 
It arises from the stochastic thermal field $\bm{H}_\mathrm{th}$, where the scalars $H_{\mathrm{th}, 1}$, $H_{\mathrm{th}, 2}$, and $H_{\mathrm{th}, 3}$ represent the components of $\bm{H}_\mathrm{th}$ projected onto the local orthonormal basis, expressed as $\bm{H}_\mathrm{th} = H_{\mathrm{th}, 1} \hat{\bm{H}}_1 + H_{\mathrm{th}, 2} \hat{\bm{H}}_2 + H_{\mathrm{th}, 3} \hat{\bm{H}}_3$. 
In this projection, the unit vectors $\hat{\bm{H}}_1$ and $\hat{\bm{H}}_2$ align parallel to the precessional torque $\bm{m}\times\bm{H}_\mathrm{eff}$ and the dissipative torque $\bm{m}\times (\bm{m} \times \bm{H}_\mathrm{eff})$, respectively, with the third basis vector defined as $\hat{\bm{H}}_3=\hat{\bm{H}}_1\times \hat{\bm{H}}_2$.
Crucially, the thermal fluctuation component $\Delta S_\mathrm{res}^{(\mathrm{th})}$ can take negative values when $\tilde{H}_\mathrm{th}^\mathrm{dmp}$ remains predominantly negative over the integration domain $(t, \bm{r})$. 
This stands in stark contrast to the deterministic component $\Delta S_\mathrm{res}^{(\mathrm{det})}$, which is independent of $\bm{H}_\mathrm{th}$ and remains strictly non-negative. 
This suggests the possibility that when the thermal fluctuation component outweighs the deterministic counterpart, the total reservoir entropy production can become temporarily negative. 
To explore this mechanism, we consider how the thermal field influences the magnetization dynamics via the projected LLG equation:
\begin{equation}
    \dot{\bm{m}} = [\dot{\bm{m}}]_\mathrm{det} + \eta \tilde{H}_\mathrm{th}^\mathrm{pre} \hat{\bm{H}}_1 - \eta \tilde{H}_\mathrm{th}^\mathrm{dmp} \hat{\bm{H}}_2,
\label{eq:sLLG_projected}
\end{equation}
where $[\dot{\bm{m}}]_\mathrm{det} = -\eta\big[\bm{m}\times \bm{H}_\mathrm{eff} + \alpha \bm{m}\times (\bm{m}\times \bm{H}_\mathrm{eff})\big]$ denotes the deterministic trajectory, and $\tilde{H}_\mathrm{th}^\mathrm{pre}=H_{\mathrm{th}, 2} + \alpha H_{\mathrm{th}, 1}$ represents the effective thermal drive for precession.
Notably, a negative $\tilde{H}_\mathrm{th}^\mathrm{dmp}$ opposes the deterministic Gilbert damping process, temporarily ``reversing'' the magnetic configuration to an earlier state. 
This backward fluctuation provides additional opportunities for the system to explore the state space, effectively increasing the attempt rate for experiencing transient negative entropy production events. 
Conversely, a positive $\tilde{H}_\mathrm{th}^\mathrm{dmp}$ accelerates the Gilbert damping process toward equilibrium. 
Therefore, even though the stochastic nature of $H_{\mathrm{th}, 1}$ and $H_{\mathrm{th}, 2}$ implies that positive and negative values of $\tilde{H}_\mathrm{th}^\mathrm{dmp}$ are equally probable \textit{a priori}, their net impact on entropy production may not simply cancel out over time.
Meanwhile, the precessional thermal drive $\tilde{H}_\mathrm{th}^\mathrm{pre}$ either drives or suppresses the precessional motion depending on its sign, further enriching the magnetization dynamics; for a detailed analysis of these coupled effects, refer to Supplementary Note~1. 
Since the average magnitudes of both $\tilde{H}_\mathrm{th}^\mathrm{dmp}$ and $\tilde{H}_\mathrm{th}^\mathrm{pre}$ scale with $\sqrt{k_\mathrm{B}T}$, such fluctuation-driven excursions become increasingly pronounced at elevated temperatures. 
Taken together, these asymmetric dynamical effects suggest a plausible physical basis for how thermal fluctuations might facilitate negative reservoir entropy production at elevated temperatures. This provides a qualitative explanation for the transient behavior observed in Fig.~\ref{fig3}(c).


The dissipated work---represented by the second term in Eq.~\eqref{eq:cost_function}---exhibits a nonmonotonic evolution during training: it initially decreases through the first 30 generations, undergoes a substantial increase near the 50th generation, and subsequently resumes a downward trend [Fig.~\ref{fig3}(a)]. This behavior suggests that minimizing dissipated work requires a training duration extending well beyond the initial convergence of the unit topological charge. Such minimization is critical for ensuring the energetic stability of the generated skyrmions. To illustrate this, we examine the magnetic energy distribution, $\rho(\mathcal{F})$, of skyrmion states generated under the same control protocol. At the 50th generation, where the dissipated work remains substantial ($\langle W_f\rangle= 458.43~\text{eV}$), $\rho(\mathcal{F})$ is significantly biased toward higher-energy regions relative to the equilibrium distribution, $\rho_{\text{eq}}(\mathcal{F})$ [Fig.~\ref{fig3}(d)]; the calculation details for $\rho_{\text{eq}}(\mathcal{F})$ are provided in Supplementary Note~7. At this stage, the average magnetic energy density $\langle \mathcal{F} \rangle$ (calculated from $\rho(\mathcal{F})$) is $1.243 \pm 0.0012~\text{MJ/m}^3$, where the error denotes the standard error of the mean. This value is notably higher than the stabilized state value of $1.189 \pm 0.0004~\text{MJ/m}^3$. By the 200th generation, however, as the dissipated work becomes minimal ($\langle W_f\rangle= 218.59~\text{eV}$), $\rho(\mathcal{F})$ closely approaches $\rho_{\text{eq}}(\mathcal{F})$, which is dominated by low-energy magnetic states [Fig.~\ref{fig3}(e)]. The corresponding average magnetic energy density, $\langle \mathcal{F} \rangle = 1.197 \pm 0.0010~\text{MJ/m}^3$, nearly converges to the stabilized state value. Our findings reveal that the convergence of the energy distribution toward the equilibrium state is gradual and consistent as the generations proceed, as illustrated in Fig.~S6. These results demonstrate that the energetic stability of skyrmions is inextricably linked to the minimization of dissipated work during the generation process.

Figure~\ref{fig3}(f) displays the evolution of the KL divergence alongside the average magnetic energy density of the skyrmions across successive DRL generations. The KL divergence exhibits a marked reduction, dropping from $D_\text{KL} \approx 0.52$ at the 50th generation to a negligible $D_\text{KL} \approx 0.03$ by the 200th generation. Given the concurrent reduction of dissipated work in this regime [Fig.~\ref{fig3}(a)], these results support our hypothesis that minimizing dissipated work facilitates the minimization of the KL divergence by lowering its upper bound. This trend is consistent with our earlier observation that $\rho(\mathcal{F})$ significantly deviates from $\rho_{\text{eq}}(\mathcal{F})$ at the 50th generation but becomes nearly indistinguishable from it by the 200th generation. Furthermore, the reduction in $D_\text{KL}$ is accompanied by a steady decline in the average magnetic energy density, which decreases from $\langle \mathcal{F} \rangle = 1.243 \pm 0.0012~\text{MJ/m}^3$ at the 50th generation to $1.197 \pm 0.0010~\text{MJ/m}^3$ by the 200th generation, asymptotically approaching the equilibrium level of $1.189 \pm 0.0004~\text{MJ/m}^3$. Consequently, these results demonstrate a robust connection between energetic stability and the KL divergence, suggesting that achieving the energetic stability of skyrmions can be accomplished via the minimization of the KL divergence.

\begin{figure*}
    \centering
    \includegraphics[width=1\linewidth]{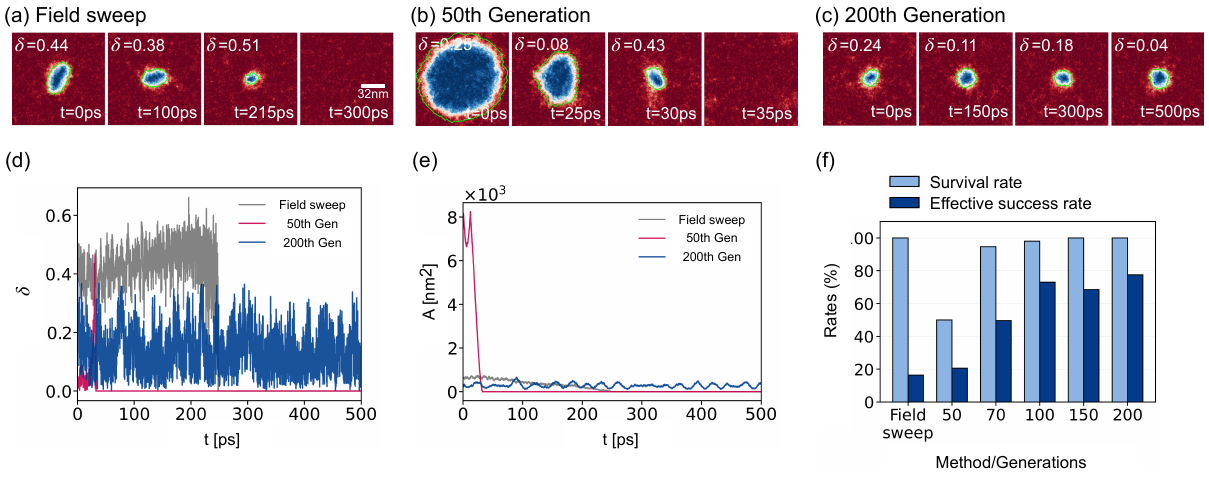}
    \caption{\textbf{Thermal relaxation dynamics and stability of skyrmions formed by different control protocols.} (a)–(c) Snapshots of representative spin configurations during relaxation under the parameter set ($T = 100$ K, $H_x = 0$, $H_z = 0.15$ T). The initial states correspond to the final configurations: (a) an antiskyrmion generated by the field-sweep protocol, (b) a skyrmion generated by the 50th-generation DRL protocol [Fig.~\ref{fig2}(b)], and (c) a skyrmion generated by the 200th-generation DRL protocol [Fig.~\ref{fig2}(c)]. The time stamp and corresponding ellipticity parameter $\delta$ are indicated in each snapshot. (d) Time evolution of $\delta$ for the trajectories shown in (a)–(c). In panels (a)--(c), the green contour marks the reference boundary used to measure the skyrmion size (area $A$; see Supplementary Note~4). (e) Time dependence of the skyrmion area $A$ for the trajectories shown in (a)–(c). (f) Survival rate and effective success rate of skyrmions evaluated over a $10$ ns relaxation period.}
    \label{fig4}
\end{figure*}

To evaluate their thermal stability, we performed relaxation simulations using the generated skyrmion textures as initial configurations. These simulations were conducted under the final parameter set of the protocols: $\mu_0\bm{H} = (0, 0, 0.15)~\text{T}$ and $T = 100$~K. 
The morphological evolution was characterized by the skyrmion area $A$ and the ellipticity $\delta$, the latter of which is defined as:
\begin{equation} \label{eq:delta}
    \delta = \frac{\lambda^+ - \lambda^-}{\lambda^+},
\end{equation}
where $\lambda^+$ and $\lambda^-$ ($\lambda^+ \geq \lambda^- > 0$) are the eigenvalues of the dissipative tensor $\mathcal{D}_{ij} = \int dx dy \frac{\partial \bm{m}}{\partial r_i} \cdot \frac{\partial \bm{m}}{\partial r_j}$ \cite{masell2020spin}, with $r_1=x$ and $r_2=y$. 
Here, $\delta = 0$ corresponds to a perfectly circular profile, while $\delta \to 1$ represents highly elliptical distortion. 
Figure~\ref{fig4}(a) depicts the annihilation process of an antiskyrmion generated via the field-sweep method. During relaxation, $A$ decreases steadily [Fig.~\ref{fig4}(e)], while $\delta$ increases from its initial value of $0.44$ to a maximum of $0.62$ at $t = 227.3~\text{ps}$ [Fig.~\ref{fig4}(d)]. 
Shortly thereafter, the antiskyrmion collapses into the FM ground state, a behavior consistent with reports that antiskyrmions are energetically unfavorable in systems with isotropic interfacial DMI \cite{koshibae2016theory, hoffmann2017antiskyrmions}. 
Similarly, Figure~\ref{fig4}(b) illustrates the annihilation of an oversized skyrmion produced by the 50th-generation protocol. Following initial breathing-mode oscillations ($t \lesssim 20~\text{ps}$), the skyrmion undergoes a rapid reduction in size [Fig.~\ref{fig4}(e)]. 
Simultaneously, elliptical deformation becomes pronounced, driving the ellipticity to a sharp peak of $\delta = 0.47$ at $t = 31~\text{ps}$ [Fig.~\ref{fig4}(d)], leading to collapse at $t = 32.7~\text{ps}$. 
One possible explanation for such annihilation is the thermally activated excitation of the elliptic distortion mode ($l=2$)~\cite{lin2014internal, desplat2018thermal}, which can induce elliptical deformation of the skyrmion prior to collapse.
By contrast, the skyrmion generated by the 200th-generation protocol remains topologically intact [Fig.~\ref{fig4}(c)], with both $\delta$ and $A$ exhibiting bounded oscillations [Fig.~\ref{fig4}(d)--(e)]; across all realizations, the ensemble-wide maxima remain bounded by $A < 428~\text{nm}^2$ and $\delta < 0.39$. This indicates that the skyrmion remains close to the equilibrium configuration and thereby reducing the likelihood of thermally activated collapse~\cite{desplat2018thermal}.

Figure~\ref{fig4}(f) summarizes the survival rate and effective success rate for each protocol. The survival rate (light bars) is defined as the fraction of skyrmions that persist throughout the $10~\text{ns}$ simulation window, while the effective success rate is defined as the fraction of simulations that ultimately yield a stable skyrmion after the full protocol and relaxation process. The field-sweep method exhibits a survival rate of $100\%$, indicating that unstable skyrmions collapse during the protocol, leaving only thermally stable configurations. In contrast, the 50th-generation protocol achieves a survival rate of only $50.0\%$, as oversized skyrmions collapse while smaller ones survive. Protocols from later training generations (100th, 150th, and 200th) show improved survival rates approaching 100\%, which are markedly higher than those of earlier generations (50th and 70th). The instability observed in early training cycles highlights the necessity of sufficient training duration to ensure the thermal stability of skyrmions. The effective success rate (dark bars) reveals a more pronounced distinction. The field-sweep method achieves an effective success rate of only $16.4\%$ due to the low probability of successful skyrmion nucleation. Similarly, the 50th-generation protocol is limited to $20.6\%$ by the instability of oversized skyrmions. By contrast, maturely trained agents (100th, 150th, and 200th generations) markedly enhance performance, with effective success rates exceeding 73.0\% and reaching 77.5\% for the 200th generation. These results demonstrate that DRL training simultaneously improves skyrmion nucleation efficiency and thermal robustness, both essential for practical skyrmion-based applications. 

While the survival rate established in Fig.~\ref{fig4}(f) discerns the ``stable'' skyrmions from the unstable ones within the 10\,ns simulation window, this analysis alone may not ensure their stability over timescales relevant for experimental observation and practical applications. To confirm thermal stability beyond this window, we performed longer relaxation simulations under identical conditions, finding that all five randomly selected skyrmions drawn from the 700 instances generated by the 200th-generation protocol survived up to 1\,\textmu s. Despite the limited sample size, this suggests that the average skyrmion lifetime likely exceeds this timescale. To further quantify the skyrmion lifetime, we estimated the annihilation energy barrier $\Delta E$ using the geodesic nudged-elastic-band method~\cite{Bessarab2015method}, yielding $\Delta E$ in the range of $7.95~k_BT$ to $21.8~k_BT$ at $T = 100$\,K depending on the unit cell size. Applying the Arrhenius law $\tau = f_0^{-1}\exp(\Delta E / k_BT)$ \cite{Bessarab2018racetrack} with a typical attempt frequency range of $f_0 = 10^9$--$10^{11}$\,Hz~\cite{Cortes-Ortuno2017-ew}, we obtain an estimated lifetime spanning from 6 to 600~\textmu s for a unit-cell volume of $1~\text{nm}^3$.
Both analyses are detailed in Supplementary Note~8. Taken together, these results confirm that the skyrmions generated by our DRL protocol exhibit lifetimes exceeding 1\,\textmu s. This is comparable to the 2.4\,\textmu s lifetime of Li-decorated FGT monolayers at the same temperature \cite{2926-8rf3}.

Finally, to assess the versatility of our DRL framework, we conducted additional training and evaluation on three distinct magnetic systems---one with the same DMI strength under open boundary conditions, and two with smaller and larger DMI strengths under periodic boundary conditions. When the 200th-generation control protocol, trained under the original conditions, was directly applied to these systems without retraining, it achieved skyrmion nucleation success rates of 57\%, 21\%, and 20\% with survival rates of 90\%, 86\%, and 100\%, demonstrating transferability across different magnetic environments. In contrast, the field sweep method achieved only a 16\% success rate and a 41\% survival rate under open boundary conditions, and failed to nucleate skyrmions in the other two cases, indicating the superior robustness of the DRL-optimized protocol. Furthermore, upon retraining the 200th-generation protocol on each target system, the success rates improved substantially to 87\%, 87\%, and 91\%, with survival rates of 72\%, 100\%, and 100\%, respectively, demonstrating the rapid adaptability of our framework to new magnetic conditions. Interestingly, the optimized field and temperature protocols consistently converged to sinusoidal forms across all systems, suggesting a possible physical origin underlying these control profiles. Refer to Supplementary Note 9 for further details.


\section{Discussion}\label{sec3}

In this work, we demonstrate that DRL provides a powerful framework for discovering optimized magnetic field and temperature control protocols to create thermally stable skyrmions in FGT. The results indicate that our physics-informed design---centered on the topological charge and the minimization of dissipated work---effectively navigates agents through complex energy landscapes to identify stable, low-energy skyrmion configurations that are otherwise difficult to access. We highlight that our strategy for minimizing dissipated work is closely linked to the principle of minimum entropy production \cite{PhysRev.96.250}. The difference between dissipated work and entropy production---defined as the magnetic energy change during the protocol (see  Supplementary Note~2 for details)---is negligible in our system. Thus, while inherently irreversible, the proposed control protocols operate close to the reversible limit, consistent with the principle of minimum entropy production.

While our specific results are directly applicable to FGT monolayers and thin films \cite{doi:10.1126/sciadv.abb5157, birch2022history, Wu2020}, this DRL framework is readily adaptable to other magnetic materials, such as Fe\textsubscript{3}GaTe\textsubscript{2} \cite{https://doi.org/10.1002/adma.202311022}, by incorporating their specific model parameters, as demonstrated by our robustness checks across various DMI strengths. 
Furthermore, the framework can be extended to bulk-DMI chiral magnets, such as MnSi \cite{Muhlbauer2009skyrmion}, FeGe \cite{Yu2011FeGe}, Cu\textsubscript{2}OSeO\textsubscript{3} \cite{Adams2012Cu2OSeO3}, and other B20 compounds \cite{Tokura2021review,Yu2010realspace}, by substituting interfacial DMI with bulk DMI. 
It is also desirable to examine various ranges of Gilbert damping coefficients beyond the currently considered low value ($\alpha=0.06$) to identify optimal operating conditions for skyrmion-based applications, as both the nucleation rate of metastable skyrmions via external perturbations such as spin-transfer torque \cite{PhysRevB.85.174416, PhysRevResearch.4.043105} and their thermal stability \cite{doi:10.1073/pnas.2122237119} are strongly dependent on $\alpha$. Even in skyrmion-based devices that do not require deterministic creation---such as Brownian-motion-based probabilistic computing \cite{Pinna2018probabilistic, Jibiki2020brownian, Zazvorka2019} or nanofluidic logic that exploits collective skyrmion flows in place of individually addressed textures \cite{doi:10.1073/pnas.2506204122}---an optimized creation protocol can substantially improve the efficiency and yield of the upstream skyrmion supply, making the framework broadly relevant.
A key insight from our study for such extensions is the dual role of the cost function components: the topological charge term enhances the nucleation success rate by leveraging thermal activation, while the dissipated work term ensures the energetic stability of the resulting skyrmions by minimizing their magnetic energy. These synergistic effects enable DRL-derived protocols to substantially outperform conventional fixed-temperature field sweeps. To further enhance the agents' capability and learning efficiency, we propose refining the cost function with additional physical descriptors, such as skyrmion area and shape anisotropy, which this study has identified as critical factors for long-term thermal stability. 


Our successful application of DRL to skyrmion nucleation highlights its potential for optimizing other generation methods---including laser pulses \cite{doi:10.1021/acs.nanolett.8b03653}, spin-transfer torque \cite{doi:10.1126/science.1240573}, and local electric field gating \cite{Hsu2017}---by leveraging the DRL agent's capability to effectively navigate the vast parameter space to identify optimal control trajectories for specific target states. More broadly, we propose that this DRL-based strategy offers a practical solution for governing the complex magnetization dynamics of both topological and non-topological spin textures. This is particularly relevant for next-generation spintronic applications where the lack of precise microscopic mechanisms and the presence of non-linear dynamics often leave the success of conventional methods to chance. By bypassing the need for an exhaustive theoretical understanding of every microscopic detail, the DRL agent autonomously discovers robust manipulation protocols that remain resilient under thermal agitation. 

\begin{acknowledgments}
J. S. and S. K. were supported by the Creation of the Quantum Information Science R\&D Ecosystem (human resources development) program through the National Research Foundation of Korea (NRF) funded by the Korean government (Ministry of Science and ICT, MSIT) (No. RS-2023-00256050) and the Brain Pool Plus Program through the NRF funded by the MSIT (No. 2020H1D3A2A03099291). K. K. was supported by an appointment to the JRG Program at the APCTP through the Science and Technology Promotion Fund and Lottery Fund of the Korean Government, the Korean Local Governments (Gyeongsangbuk-do Province and Pohang City), and the NRF funded by the MSIT (No. RS-2026-25499525).
\end{acknowledgments}

\bibliography{ref}
\end{document}